# Convection roll-driven generation of supra-wavelength periodic surface structures on dielectrics upon irradiation with femtosecond pulsed lasers

George D.Tsibidis, [1♣] Evangelos Skoulas, [1,2] Antonis Papadopoulos, [1,3] and Emmanuel Stratakis [1,3]*

[1] *Institute of Electronic Structure and Laser (IESL), Foundation for Research and Technology (FORTH), N. Plastira 100, Vassilika Vouton, 70013, Heraklion, Crete, Greece*
[2] *VEIC, Department of Ophthalmology, School of Medicine, University of Crete, Greece*
[3] *Materials Science and Technology Department, University of Crete, 71003 Heraklion, Greece*

The significance of the magnitude of Prandtl number of a fluid in the propagation direction of induced convection rolls is elucidated. Specifically, we report on the physical mechanism to account for the formation and orientation of previously unexplored supra-wavelength periodic surface structures in dielectrics, following melting and subsequent capillary effects induced upon irradiation with ultrashort laser pulses. Counterintuitively, it is found that such structures exhibit periodicities, which are markedly, even multiple times, higher than the laser excitation wavelength. It turns out that the extent to which the hydrothermal waves relax depends upon the laser beam energy, produced electron densities upon excitation with femtosecond pulsed lasers, magnitude of the induced initial local roll disturbances and the magnitude of the Prandtl number with direct consequences on the orientation and size of the induced structures. It is envisaged that this elucidation may be useful for the interpretation of similar, albeit large-scale periodic or quasi-periodic structures formed in other natural systems due to thermal gradients, while it can also be of great importance for potential applications in biomimetics.

**PACS:** 78.20.Bh 64.70.D- 42.65.Re 77.90.+k

The predominant role of convective flow in nonequilibrium spatial pattern formation has been demonstrated in various phenomena in nature such as on Earth's land and sea, as well as on planet's surface when large thermal or wind speed gradients are developed [1-4]. Similar fluid instabilities are also encountered in many industrial applications: heat exchangers [5], evaporative cooling devices, and chemical vapor process [6], film flow in inclined porous substrates used in oil pipes [7], manufacturing of high purity semiconductor crystals [8].

Similar patterns and more specifically, periodical structure formation are also induced on the surface or volume of many solids upon irradiation with laser beams [9-11]. This modification usually requires a solid to liquid phase transition followed by fluid movement and capillary effects. Whether classical Navier-Stokes equations and how a Newtonian fluid mechanics could determine quantitatively the characteristics of the molten material dynamics, still remain an open question.

The formation of surface and bulk periodic structures gives rise to unique material properties and it has received considerable attention over the past decades due to a wide scope of applications regarding micro/nano-structuring of materials [12]. In this context, one timely area of exploration has been the ultrashort pulsed laser induced periodic structuring of dielectrics, due to its applicability in telecommunications, biomedicine and biomimetics [13-17]. The predominant physical mechanism to explain the formation of laser induced periodic surface structures (LIPSS) suggests that excited electron densities modify the dielectric constant and the refractive index $n$ [18] leading to low spatial frequency LIPSS (LSFL). Alternative mechanisms to explain a spontaneous pattern formation (self-organisation) in solids include the analogy with ion-beam-sputtering [19]. These experimentally observed structures are oriented either parallel [20-23] or perpendicularly [20, 24] to the polarisation of the laser beam and possess sub- (or near) wavelength periodicities. Previous experimental studies showed that development of structures with larger periodicity than that of the laser beam wavelength (*supra*-wavelength) is also possible while they are oriented parallel to the beam polarisation [25]. However, the current theoretical framework fails to predict the formation of *supra*-wavelength structures. Although Marangoni convection has been presented to interpret the formation of *sub*-wavelength ripples [26], a different mechanism based on the physics of convection roll movement are considered to explain the creation of *larger* structures as well as their orientation.

In this Rapid Communication, we present the physical mechanism that accounts for the formation of *supra*-wavelength structures on dielectrics, following irradiation with high photon energy femtosecond laser pulses. Hence, we focus on a spatiotemporal pattern formation in a physical system that is driven away from equilibrium by intensive heating. We especially focus on fused silica, however similar structures have been observed in other dielectrics and different materials [27].

As shown in Fig. 1a, two types of parallel structures of distinctly different size ($λ_1 > λ_2$) are produced on the surface of silica as derived using a Fast Fourier Transform (FFT) (Fig.1b,c,d,e). We show that the creation of, low spatial frequency, microstructures cannot be interpreted through the commonly used electrodynamic formulations [20]. On the contrary, this work reveals that convection roll-driven hydrodynamic phenomena in the melted glass are behind their formation. Specifically, to account for the phase transitions and subsequently induced modifications in the melted profile, we introduce a theoretical framework based on an electron excitation (induced by the laser impact) and electron-lattice heat diffusion model coupled with Navier-Stokes equations [28]. The hydrodynamic component describes the molten material dynamics and resolidification, assuming the material as an incompressible Newtonian fluid and considering hydrothermal convection as the predominant physical process that gives rise to periodic structures' formation [29]. Simulations were performed using a *p*-polarised laser beam of fluence $E_d$=3.33J/cm$^2$, pulse duration $\tau_p$=170fs, Gaussian beam radius $R_0$=15μm, for two wavelengths $\lambda_L$=513nm and 1026 nm. Laser irradiation with one pulse ($NP$=1) produces solely a shallow crater due to Marangoni convection as reported also in other solids [28, 30]. It is notable that prior to the investigation of the surface tension driven fluid movements, a small mass removal has been assumed as a result of intense heating followed by an associated recoil pressure. Mass removal is observed experimentally while theoretically is manifested to occur for material temperatures well above the boiling point $T_b$ (~3223K) [31]. A similar approach has been considered to explore approaches to simulate minimal mass ejection in semiconductors [28]. Due to laser irradiation conditions that lead to transition to a liquid phase (i.e. lattice temperature higher than the melting point temperature $T_m$=1988K), the movement of the $T_m$ isothermal line is used to describe dynamics of the molten profile and the induced surface morphology after resolidification.

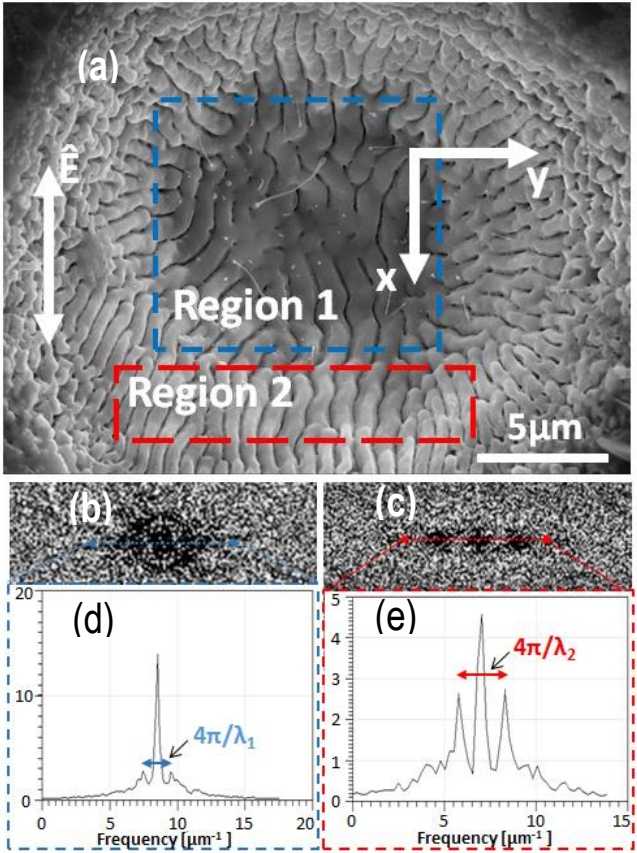

FIG. 1. (Color online) (a) Morphological changes induced on $SiO_2$ following irradiation. ($NP$=30, $E_d$=3.33/cm$^2$, $\lambda_L$=513nm, $\tau_p$=170fs, the double-ended arrow indicate the laser beam polarization). (b), (c) Periodicities calculated using a FFT on regions 1 and 2, respectively, while (d) and (e) are the relevant intensity profile across the crossline in the frequency domain ($\lambda_1 > \lambda_2$).

To explore surface modification with increasing number of pulses, the inhomogeneous energy deposition into the irradiated material is computed through the calculation of the efficacy factor, $\eta$, as described in the Sipe-Drude model [32]. This parameter describes the efficacy with which the surface roughness induces inhomogeneous radiation absorption. It is closely related to the electron density, $n_e$, and therefore it influences the frequency of possible periodical structures [20, 27, 32]. To simulate the excitation mechanism in fused silica, we consider a revised version [27] of the multiple rate equation model [33] to accommodate both self-trapped exciton (STE) states and free electron relaxation. It is evident that for $\lambda_L$ =513nm (Fig.2a and [27]) and $NP$=3, $n_e$ values inside the region of $r_G$≤8μm do not lead to parallel periodic structure formation as determined from the computation of $\eta$ for $4.4\times10^{21}$cm$^{-3}$≤$n_e$≤$7.5\times10^{21}$cm$^{-3}$ where normalized wave vector components and fluence equal to damage threshold were considered. More specifically, periodic structures should be developed with periodicity determined by the point where the efficacy factor exhibits sharp points. It is evident that at these electron densities, this is not possible which suggests ripples are not formed as the efficacy factor along $k_y$ (i.e. normalized wavevector $|\mathbf{k}_L|=\lambda_L/\Lambda$, where $\Lambda$ stands for the periodicity) does not yield sharp points (Fig.2b) [27]. This is illustrated in (Fig.2b) where the only experimentally possible ripple orientation (parallel to the beam polarisation) for these range of electron densities is not promoted by the calculated efficacy factor value [20]. By contrast, at $r_G$>8μm, $n_e$ is smaller and LIPSS are formed as the efficacy factor values allow their formation [27]. On the other hand, an increase in the surface roughness with increasing irradiation is capable to modify the picture and lead to periodical structure formation even inside the $r_G$≤8μm region. In that region, a decreased electron density is gradually developed by increasing the number of pulses. This results from the increase of the surface profile curvature (i.e. due to minimal material removal and to mass displacement because of surface tension gradients) that leads to smaller energy deposition and electron excitation.

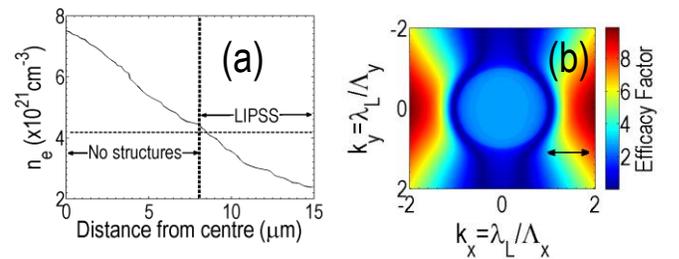

FIG. 2. (Color online) (a) Spatial distribution of electron density along *y*-direction (horizontal *dotted* line defines the electron density threshold for periodic structure formation) for $NP$=2 ($\lambda_L$ =513nm). (b) Efficacy factor for $n_e$=7.8×10$^{21}$cm$^{-3}$ at $x=y=0$. ($E_d$=3.33/cm$^2$, $\lambda_L$=513nm, $\tau_p$=170fs, $NP$=3). (the double-ended arrow indicates the laser beam polarization, $\lambda_L$=513nm).

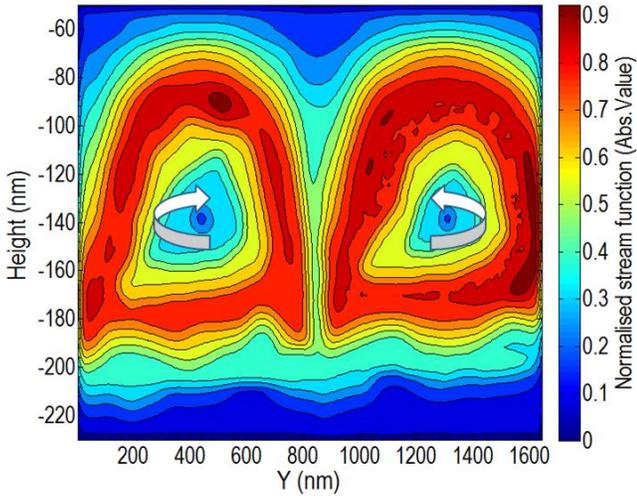

FIG. 3. (Color online) Stream function field at $t$=2ns along the Y-axis (*white* arrows indicate the fluid movement direction considering the absolute value of the stream function normalised to 1). ($E_d$=3.33/cm$^2$, $\lambda_L$=513nm, $\tau_p$=170fs, $NP$=12).

In order to describe the morphological profile that is produced as a result of phase transition, fluid transport, and solidification process, a numerical solution of the Navier-Stokes equations is followed using a finite difference scheme in a staggered grid [28, 34]. Stress-free and no-slip boundary conditions are imposed on the liquid-solid interface while shear stress boundary conditions are assumed on the free surface. Although surface modification occurs both inside and outside the region $r_G$≤8μm and hydrodynamics is explored in both regions, as explained in the introduction, LSFL ripples are generated when electron densities are relatively small [20]. The underlying electrodynamics-related processes precede the phase transition and they determine the ripple size upon solidification. By contrast, inside $r_G$≤8μm, irradiation produces large electron densities and LSFL ripples do not occur and therefore surface modification is purely determined by lattice cooling processes (i.e. fluid dynamics and resolidification). Using a common approach, equations describing the overall convective flow (i.e. base flow) are used followed by a linear stability analysis of small perturbations to the flow [5, 7, 27, 35] to describe melt dynamics in a curved profile. Normal mode analysis requires the introduction of a $\vec{k}_R$ vector that characterizes the wavenumber of the induced counter-rolls. To elaborate on the melt hydrodynamic flow and predict the generation of hydrothermal waves in thin liquid films through heat transfer and Marangoni convection, the conditions that lead to convection roll structures are evaluated [7, 29, 36-38]. In principle, the Marangoni number, $M$, that represents the ratio of rate of convection and rate of conduction, is used to characterize the flow due to surface tension gradients through the relation $M=(\partial\sigma/\partial T)\Delta T\Delta H/(\alpha\mu)$, where $\partial\sigma/\partial T$, $\Delta T$, $\Delta H$, $\alpha$, and $\mu$ are the surface tension gradient, average temperature difference between the lower and upper layer of the liquid, average thickness of the fluid, thermal diffusivity and dynamic viscosity of the molten material. Stability of the counter-rolls occur at the minimum of the Marangoni function and therefore for $NP$=7, $|\vec{k}_R|/(2\pi)$ ~569nm (for $NP$<7, no stability occurs for $\lambda_L$ =513nm) [27]. The dynamics of the convection rolls (at time $t$=2ns and for $NP$=12) is illustrated in Fig.3 where a decreased vorticity is observed inside the roll, while the stream function at the bottom of the liquid indicates an almost parallel flow along Y-axis (i.e. stream function $\psi$ is computed by solving the Poisson equation $\vec{\nabla}^2\psi = \vec{\nabla}\times\vec{U}$, where $\vec{U}$ is the velocity at every point). Due to the opposite direction of rotation of the two rolls, the vorticity $\vec{\nabla}\times\vec{U}$ drops significantly at the meeting point of the two waves. Note that the *absolute* value of the stream function field for the convection roll on the right was considered (negative values is demonstrated by a counterclockwise roll rotation).

It is evident that dynamics of the molten profile is closely related to the laser polarization. As a result of the inhomogeneous heating, the produced temperature field induce surface stresses that drive a shear flow of the molten layer *parallel* to the thermal gradient [29] (Fig.3). It appears that thermal gradients are more enhanced in a plane that is perpendicular to the plane of incidence and therefore the preferred direction of axis of rotation of convection rolls is *parallel* to the laser polarisation which is also evident in other physical systems [1-4]. This thermal-convective instability is indicative for fluids characterised by a large Prandtl number $P_r$ (i.e. $P_r$>10$^4$ for molten silica) leading to *longitudinal* counter-rotating rolls [27, 29, 35, 38] in contrast to *transverse* convection rolls that occur for liquids with low Prandtl number [34, 39]. We, hence, postulate that when the temperature difference within the melt exceeds a critical value, convection rolls generated by surface tension gradients and hydrothermal waves will induce surface tractions, developed *parallel* to ripples orientation but with an increased periodicity.

The convection roll dynamics followed by an analysis of stability conditions suggest that the initially free of periodic structures space (i.e. inside the $r_G$ region (Fig.4a) due to an enhanced $n_e$) upon melting will produce *supra*-wavelength sized and constrained periodic structure formation (see Fig.4b and [27]). Dynamics of the molten material is followed through the solution of the Navier-Stokes equation where the morphology of the surface profile is determined by the condition that resolidification occurs when the temperature isothermal falls below $T_m$. By contrast, a cohabitation of both sub-to-wavelength ripples and supra-wavelength sized periodical structures after resolidification is illustrated in Fig.4b.

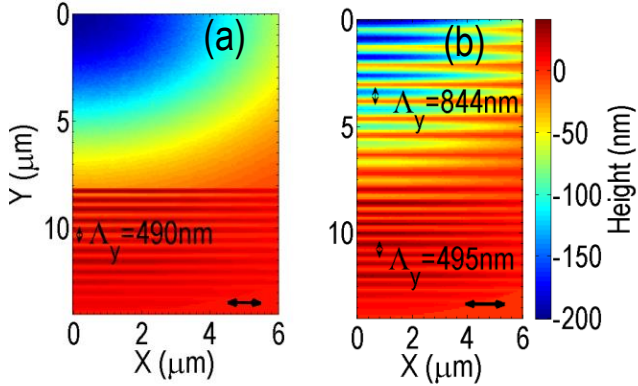

FIG. 4. (Color online) Upper view of surface profile: (a) *NP*=6. (b) *NP*=12. ($E_d$=3.33J/cm$^2$, $\lambda_L$=513nm, $\tau_p$=170fs, the double-ended arrow indicates the laser beam polarization).

To validate the theoretical results, an experimental protocol using commercial polished samples of SiO$_2$ of 99.9% purity and average thickness 1mm. An Yb:KGW laser source was used to produce linearly polarized pulses of with pulse duration set as 170fs, 60 KHz repetition rate and 513nm and 1026nm central wavelength. Results indicate that groove periodicity increases both on the fluence (Fig.5a) and number of pulses (Fig.5b). Similar behaviour has been observed experimentally for grooves in silicon while on the other hand, ripples with periodicity smaller than $\lambda_L$ (~$\lambda_L/n$ or ~ $\lambda_L$) was essentially independent of the fluence [40]. In SiO$_2$, an increase of the groove periodicity with number of pulses is also related to incubation effects that lead to an enhanced excitation for the applied fluence and larger periodicity for the induced counter rolls. Although this is not possible for irradiation with one pulse, incubation effects and STE states will cause reduction of the fluence damage threshold at about 1/3 of the value for *NP*=1 [41]. A similar enhancement of the hydrodynamic effect is responsible at higher fluences. It is important to underline the influence $\lambda_L$ in the resulting groove periodicity. As emphasized above, the initial production of *supra*-wavelength sized grooves in the $r<r_G$ region for $\lambda_L$=513nm is reflected in the graph where accumulation effects through repetitive irradiation (or fluence increase) will enhance the hydrodynamical stresses by increasing the groove periodicity substantially. By contrast, for $\lambda_L$=1026nm, this effect does not occur firstly and therefore only conventional sub-(to near-) wavelength sized rippled structures are initially formed [20] as electron densities does not lead to ripple depletion. Nevertheless, for larger values of fluence or *NP*, grooves are produced even for $\lambda_L$=1026nm which is based on the proposed hydrodynamics originated mism. More specifically, for fluences in the range 2.9-3.2J/cm$^2$, there is no formation of periodic structures as $n_e$ is large while for higher values, hydrodynamics takes over and leads gradually to groove formation (Fig.5a). The fluence upper limit for the simulation for $\lambda_L$=1026nm is set to 4.33J/cm$^2$ as ablation occurs for higher values and in that case, the proposed mechanism has to be modified properly. It is also evident that low fluences (~1.2J/cm$^2$) lead to surface modification and therefore material damage for *NP*=20 (Fig.5a). A similar transition from ripples to grooves occurs with increasing *NP*; while rippled structures are produced for low *NP*, grooves are developed for *NP*>18. It is evident at *NP*=15 cease to develop while for *NP*=18, grooves are formed due to hydrodynamics. We note that the monotonicy of the ripple periodicity with increasing number of pulses differs from that demonstrated in metals [30, 42] or semiconductors [28, 43]. This observation supports the importance of incubation effects in the ripple formation on dielectrics. Such effects lead to enhanced absorption resulting, for subsequent laser pulses, to larger concentration of excited carriers in the conduction band. This monotonicity is then explained by a predicted larger ripple periodicity with increasing $n_e$ (i.e. by computing the efficacy factor and the produced ripple periodicity for various $n_e$) [20, 41]. On the other hand, a comparison of the groove formation with increasing *NP* in silicon (characterized by a *low* Prandtl number) and dielectrics (with a *high* Prandtl number) shows that the mechanisms are different: in semiconductors, LIPSS are formed first due to surface plasmon excitation (SPP), then excitation is suppressed before further irradiation gives rise to convection rolls *perpendicularly* to LIPSS as fluid transport occurs along the wells of the ripples [34]; by contrast, in dielectrics, SPP are not excited, and hydrothermal waves develop *parallel* to the polarisation or parallel to ripples (provided that ripples are initially formed, as seen in Fig.5).

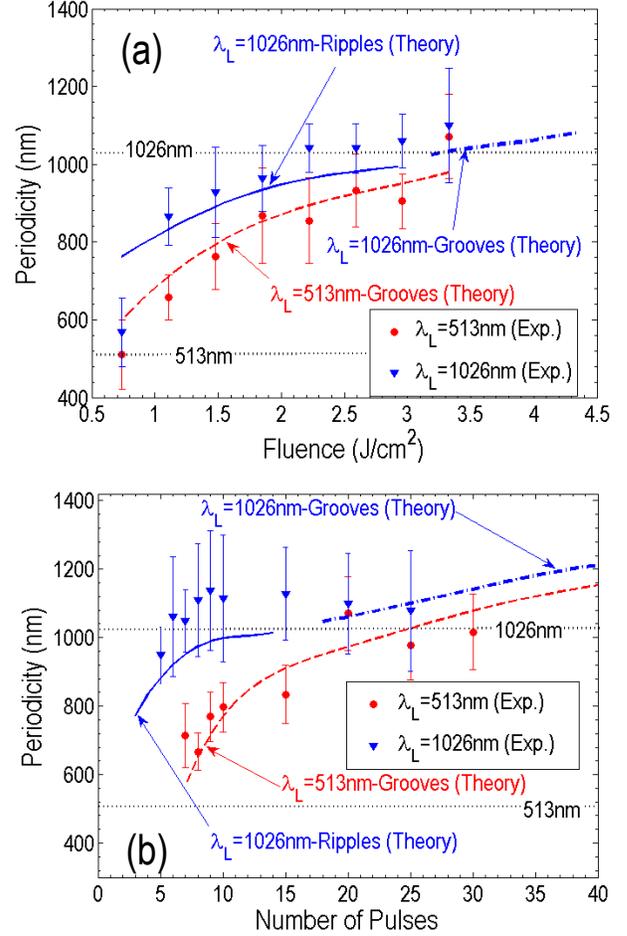

FIG. 5. (Color online) Theory *vs* Experiment: (a) Fluence dependence of periodic structures for *NP*=20. (b) Periodic structure dependence on *NP* for $E_d$=3.33/cm$^2$.

In conclusion, the physical mechanism to explain the previously unexplored produced supra-wavelength periodic surface structure formation on dielectrics was presented. The role of hydrodynamic instability in convection roll-driven formation of these large structures was particularly elucidated. Apart from the understanding of the physical mechanism behind the surface modification in dielectrics, the present work aims to offer a broader scientific benefit in diverse fields. A deeper insight into problems related to instabilities, including self-organisation, complex spatiotemporal behavior, and turbulence are expected to enhance the fundamental knowledge in pattern forming systems encountered in physics, nonlinear optics, chemistry and biology [44], with immediate fundamental and industrial merit.

This work was funded by the EU-funded project *LinaBiofluid* (665337)

♣ tsibidis@iesl.forth.gr
* stratak@iesl.forth.gr

# Supporting Material for 'Convection roll-driven generation of supra-wavelength periodic surface structures on dielectrics upon irradiation with femtosecond pulsed lasers'

George D.Tsibidis, [1,♣] Evangelos Skoulas, [1,2] Antonis Papadopoulos, [1,3] and Emmanuel Stratakis [1,3*]

[1] *Institute of Electronic Structure and Laser (IESL), Foundation for Research and Technology (FORTH), N. Plastira 100,*
*Vassilika Vouton, 70013, Heraklion, Crete, Greece*
[2] *VEIC, Department of Ophthalmology, School of Medicine, University of Crete, Greece*
[3] *Materials Science and Technology Department, University of Crete, 71003 Heraklion, Greece*

### A. Electron excitation and energy absorption.

The absorption of light in transparent materials must be nonlinear as a single photon does not have enough energy to excite electrons from the valence to the conduction band. Following nonlinear photo-ionisation (where both multiphoton and tunnelling ionisation can occur in different regimes depending on the laser intensity), avalanche ionisation and formation of self trapped excitons (STE) lead to a variation of the electron densities. Due to the large energy deposition with employment of femtosecond pulses, photo-ionisation is possible. To calculate the laser absorption, the densities of the excited electron and the STE are computed by solving simultaneously the following set of equations based on a revised version of the the Multiple Rate Equation model (MRE) [40]

$$\begin{cases} \partial_t n_1 = \frac{n_v - n_e - n_s}{n_v} PI(E_G^{(1)}) + 2\tilde{\alpha} n_k - W_{1pt} n_1 + \frac{n_s}{n_v} PA(E_G^{(2)}) - \frac{n_1}{\tau_r} \\ \partial_t n_j = W_{1pt} n_{j-1} - W_{1pt} n_j - \frac{n_j}{\tau_r}, 1 < j < k, \\ \partial_t n_k = W_{1pt} n_{k-1} - \tilde{\alpha} n_k - \frac{n_k}{\tau_r} \\ \partial_t n_s = -\frac{n_s}{n_v} PA(E_G^{(2)}) + \frac{n_e}{\tau_r} \end{cases} \quad (1)$$

where $n_e$, $n_v$, $n_s$ are the number densities of the excited electrons, STE states, and valence band electrons ($n_v$=2.2×10$^{22}$cm$^{-3}$), respectively. We note that a more accurate approach requires the use of multiple rate equations to consider that the most energetic electrons in the conduction band induce collisional excitation [33]. The last term in the first equation corresponds to free electron decay that is characterised by a time constant $\tau_r$ ($\tau_r$~150fs in fused silica) that leads to a decrease of the electron density. The photoionsation rates are computed using the Keldysh formulation [45]

$$PI = \frac{2\omega_L}{9\pi} \left(\frac{m_r \omega_L}{\gamma_2 \hbar}\right)^{3/2} \times \Theta(\gamma, x) \exp\left[-\pi \langle x+1 \rangle \frac{K(\gamma_2) - E(\gamma_2)}{E(\gamma_1)}\right] \quad (2)$$

where $\gamma = \omega_L \sqrt{m_r E_G^{(i)}} / (e|\vec{E}|)$ is the Keldysh parameter for the band-gap $E_G^{(i)}$ ($i$=1,2) which is dependent on the electron charge $e$, the frequency $\omega_L$ and the field $|\vec{E}|$ of the laser beam, the electron reduced mass $m_r$=0.5$m_e$ and $\gamma_2 = \gamma / \sqrt{1+\gamma^2}$ and $\gamma_1 = \gamma_2 / \gamma$. Also, the density of the excited electrons is $n_e = \sum_{j=1}^{k} n_j$. Furthermore, $\langle x+1 \rangle$ stands for the integer part of the number $x+1$, where $x=2E_G^{(i)} E(\gamma_1) / (\pi \gamma_2 \hbar \omega_L)$ while $K$ and $E$ are the complete elliptic integrals of the first and second kind, respectively. Also,

$$\Theta(\gamma, x) = \sqrt{\frac{\pi}{2K(\gamma_1)}} \sum_{N=0}^{\infty} \exp\left[-N\pi \frac{K(\gamma_2) - E(\gamma_2)}{E(\gamma_1)}\right] \times \Phi\left[\sqrt{\frac{\pi^2 (2\langle x+1 \rangle - 2x + N)}{K(\gamma_1) E(\gamma_1)}}\right] \quad (3)$$

where $\Phi(z) = \int_0^z exp(y^2-z^2)dy$. Finally, the avalanche ionisation rate $A$ is given by the following expression [46, 47]

$$A(E_G^{(i)}) = \frac{e^2\tau_c I}{c\varepsilon_0 nm_r\left(\omega_L^2(\tau_c)^2+1\right)\varepsilon_{crit}} = \frac{e^2\tau_c I}{c\varepsilon_0 nm_r\left(\omega_L^2(\tau_c)^2+1\right)(1+m_r/m_e)\left(E_G^{(i)}+e^2|\vec{E}|^2/(4m_r\omega_L^2)\right)} \quad (4)$$

where $\varepsilon_{crit}$ stands for the minimum impact ionisation energy, $c$ is the speed of light, $\varepsilon_0$ stands for the vacuum permittivity, $n$ is the refractive index of the material, while $I$ is the intensity and $\tau_c$ is the electron collision time ($\tau_c$~0.5fs [20]). According to the MRE model, the parameters $\tilde{\alpha}$ and $W_{1pt}$ which denote the impact ionization (the asymptotic value of $\tilde{\alpha}$ is assumed) and one-photon absorption probabilities, respectively, are equal to [40]

$$\begin{aligned}\tilde{\alpha} &= \left(\left|\sqrt[k]{2}\right|-1\right)W_{1pt} \\ W_{1pt} &= \frac{A(E_G^{(1)})}{\log(2)\left(\left|\sqrt[k]{2}\right|-1\right)}\end{aligned} \quad (5)$$

We note that in Eq. (5) the superscript $k$ obtains the following values: $k(\lambda_L=513nm)=7$ and $k(\lambda_L=1026nm)=15$ at $E_d=3.33J/cm^2$.

With respect to the intensity of the laser beam $I$, previous studies consider that the attenuation of the local laser intensity is determined by multiphoton ionisation and inverse bremsstrahlung absorption [48-50]. Nevertheless, the presence of STE states and the possibility of retransfer of them to the conduction band through multiphoton ionisation require modification of the spatial intensity distribution and therefore in a revised model it should read

$$\frac{\partial I(t,\vec{x})}{\partial z} = -N_{ph}^{(1)}\hbar\omega\frac{n_v-n_e-n_s}{n_v}PI(E_G^{(1)}) - \alpha(n_e)I(t,\vec{x}) - N_{ph}^{(2)}\hbar\omega\frac{n_s}{n_v}PI(E_G^{(2)})$$

$$I(t,x,y,z=0) = (1-R(t,x,y,z=0))\frac{2\sqrt{\log(2)}}{\sqrt{\pi}\tau_p}E_d \exp\left(-4\log(2)\left(\frac{t-3\tau_p}{\tau_p}\right)^2\right)\exp\left(-\frac{x^2+y^2}{(R_0)^2}\right) \quad (6)$$

where $N_{ph}^{(i)}$ corresponds to the minimum number of photons necessary to be absorbed by an electron that is in the valence band ($i$=1) or the band where the STE states reside ($i$=2) to overcome the relevant energy gap and reach the conduction band. Also, $R(t,x,y,z=0)$ stands for the reflectivity of the material, $E_d$ is the fluence while $R_0$ is the irradiation spot radius ($R_0$=15μm in our simulations).

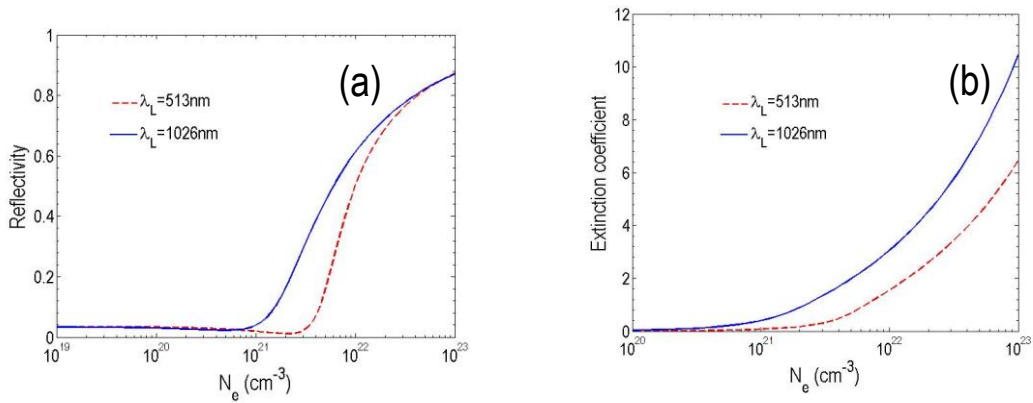

FIG. 1 Theoretical computation of the electron density dependence of reflectivity (a) and extinction coefficient (b) after irradiation with $\lambda_L$=513nm and $\lambda_L$=1026nm.

The transient optical properties of the irradiated material (i.e. refractive index, absorption coefficient and reflectivity) are computed through the evaluation of the dielectric constant for materials with a band gap $E_G^{(1)}$ [18]

$$\varepsilon(n_e) = 1 + (E_G^{(1)} - 1)\left(1 - \frac{n_e}{n_v}\right) - \frac{n_e e^2}{\varepsilon_0 m_r \omega_L^2} \frac{1}{1 + i\frac{1}{\omega_L \tau_c}} \tag{7}$$

Refractive index and absorption coefficient computation yields the electron density dependence that is illustrated in Fig.1, respectively. Furthermore, the electron density evolution is sketched in Fig.2 for $E_p$=3.33/cm$^2$. The *red* line defines the optical breakdown (critical density $n_{cr}$=4.25×10$^{21}$cm$^{-3}$ for $\lambda_L$=513nm). We notice that the produced electron density exceeds the critical density that is sufficient for optical breakdown ($n_{cr}$=(2πc/(λ$_L$e))$^2$m$_e$ε$_0$, where the plasma frequency is equal to the laser frequency). Despite the conventional methodology to correlate this value with material damage, we proceed with an alternative strategy. More specifically, we explore the electron-lattice relaxation processes and define the onset of the surface modification when the lattice temperature exceeds the melting point temperature of the material ($T_m$=1738K). Hence, we consider that surface morphology is initiated when the relaxation dynamics produces material malting. On the other hand, ablation is assumed to be related to the condition at which the lattice temperature of the material reaches 0.9$T_{cr}$ ($T_{cr}$ stands for the thermodynamical critical value which, in the case of SiO$_2$, equals 6303K).

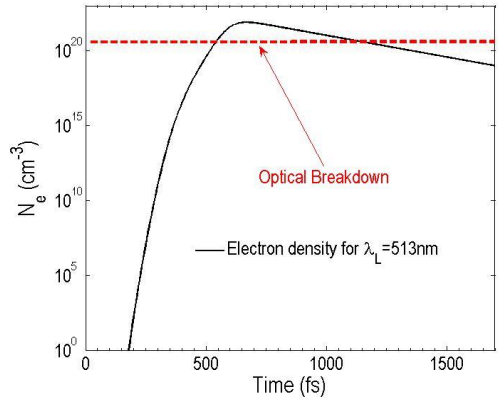

FIG. 2. Electron density evolution ($\lambda_L$=513nm, $\tau_p$=170fs, $E_d$=3.33/cm$^2$).

### B. Electron-Lattice relaxation processes.

Due to the metallic character of the excited material, a TTM model can describe the spatio-temporal dependence of the temperatures $T_e$ and $T_L$ of the electron and lattice subsystems [27, 48, 51, 52], respectively

$$C_e \frac{\partial T_e}{\partial t} = \vec{\nabla}\left(k_e \vec{\nabla} T_e\right) - g\left(T_e - T_L\right) + S(\vec{x}, t)$$
$$C_L \frac{\partial T_L}{\partial t} = g\left(T_e - T_L\right) \tag{8}$$

The source term has been modified properly to take into account all quantities that contribute to the total electron energy balance which also changes by the transient variation of the electron density. Hence, the complete expression for source term $S(\vec{x},t)$ is given by

$$S(\vec{x},t) = \left(N_{ph}^{(1)} \hbar\omega - E_G^{(1)}\right) \frac{n_v - n_e - n_s}{n_v} PI(E_G^{(1)}) - E_G^{(1)} \tilde{\alpha} \frac{n_k}{n_V} + \alpha(n_e) I(t, \vec{x}) - \frac{3}{2} k_B T_e \frac{\partial n_e}{\partial t}$$
$$+ \left(N_{ph}^{(2)} \hbar\omega - E_G^{(2)}\right) \frac{n_s}{n_v} PI(E_G^{(2)}) - \frac{3}{2} k_B T_e \frac{n_e}{\tau_r} \tag{9}$$

to account for photoionisation of electrons (both from the valence and the STE states), avalanche ionisation, free electron absorption and energy loss due to trapping processes. We point out that a term that describes the divergence of the current of the carriers ($\vec{\nabla}\cdot\vec{J}$) has not been taken into account (it appears when investigation of carried dynamics in silicon upon irradiation with ultrashort pulses [53], however, it turns out that for small pulse durations its contribution is negligible [54]). The temperature, electron density and temporal dependence of the thermophysical properties, $C_e$, $k_e$ are provided from well-established expressions derived from free electron gas based on the metallic character of the excited material (see, for example, [48, 55] for explanation of the terminology)

$$
\begin{aligned}
&E_F = \frac{(hc)^2}{8m_e c^2}\left(\frac{3}{\pi}\right)^{2/3}(n_e)^{2/3} \\
&g(\varepsilon) = \frac{8\sqrt{2}}{h^3 (m_e)^{2/3}}\sqrt{\varepsilon} \\
&\mu(n_e,T_e) = E_F\left[1 - \frac{\pi^2}{12}\left(\frac{k_B T_e}{E_F}\right)^2 + \frac{\pi^2}{80}\left(\frac{k_B T_e}{E_F}\right)^4\right] \\
&\langle\varepsilon\rangle = \frac{\int_0^\infty \exp\left(-\left((\varepsilon-\mu)/(k_B T_e)+1\right)\right)g(\varepsilon)\varepsilon d\varepsilon}{\int_0^\infty \exp\left(-\left((\varepsilon-\mu)/(k_B T_e)+1\right)\right)g(\varepsilon)d\varepsilon} \\
&C_e(n_e,T_e) = n_e \frac{\partial\langle\varepsilon\rangle}{\partial T_e} \\
&k_e(n_e,T_e) = \frac{1}{3}(u_e)^2 \tau_c C_e(n_e,T_e)
\end{aligned}
\qquad(10)
$$

On the other hand, $C_L$=1.6 J/(cm$^3$K) [55] while the coupling constant is estimated to be $g=g_0(n_e)^{2/3}$, where $g_0$=0.6×10$^{-1}$W/(mK) [48].

### C. Fluid dynamics and formation of convection rolls.

In this section we will provide the fundamentals of the fluid mechanics mechanisms that lead to formation of hydrothermal waves based on previous works(see [29, 35, 36, 38, 39] and references therein) considering also a revision due to the inclined profile resulted from the ripple geometry [7]. The molten material parameters for fused silica are the following: viscosity~100N/m$^2$, surface tension $\sigma$=0.310N/m, thermal expansion coefficient $\alpha$=3.65×10$^{-9}$/K, and density $\rho$~2.050gr/cm$^3$.

We consider an incompressible fluid with small variations $\Delta T$ about a constant value $T_0$ leading to variations in the local fluid density (Boussinesq approximation that assume buoyancy forces)

$$
\begin{aligned}
T &= T_0 + \Delta T \\
\rho &= \rho_0 - \alpha\rho_0\Delta T
\end{aligned}
\qquad(11)
$$

where $\alpha$ is the coefficient of thermal expansion of the material and $\rho$ is taken constant ($\rho=\rho_0$) everywhere except the body force in the following equation

$$
\rho_0\left(\frac{\partial\vec{u}}{\partial t} + \vec{u}\cdot\vec{\nabla}\vec{u}\right) = \vec{\nabla}\cdot\left(-P\mathbf{1} + \mu(\vec{\nabla}\vec{u}) + \mu(\vec{\nabla}\vec{u})^T\right) + \rho_0(1-\alpha\Delta T)\vec{g}
\qquad(12)
$$

where $\vec{g}$ is the acceleration of gravity. Vapour ejected creates a back (recoil) pressure on the liquid free surface which in turn pushes the melt away in the radial direction. The recoil pressure and the surface temperature are usually related according to the equation [56]

$$P_r = 0.54 P_0 \exp\left( L_v \frac{T_L^S - T_b}{R T_L^S T_b} \right) \qquad (13)$$

where $P_0$ is the atmospheric pressure (i.e. equal to $10^5$ Pa), $L_v$ is the latent heat of evaporation of the liquid, $R$ is the universal gas constant, and $T_L^s$ corresponds to the surface temperature, and $T_b$ is the boiling temperature (~3223K). Given the radial dependence of the laser beam, temperature decreases as the distance from the centre of the beam increases; at the same time, the surface tension in pure dielectric decreases with growing melt temperature (i.e $d\sigma/dT<0$), which causes an additional depression of the surface of the liquid closer to the maximum value of the beam while it rises elsewhere. Hence, spatial surface tension variation induces stresses on the free surface and therefore a capillary fluid convection is produced. Moreover, a precise estimate of the molten material behaviour requires a contribution from the surface tension related pressure, $P_\sigma$, which is influenced by the surface curvature and is expressed as $P_\sigma = K\sigma$, where $K$ is the free surface curvature. The role of the pressure related to surface tension is to drive the displaced molten material towards the centre of the melt and restore the morphology to the original flat surface. Thus, pressure equilibrium on the material surface implies that the pressure in Eq.12 should outweigh the accumulative effect of $P_r + P_\sigma$.

In a simplified scenario in which temperature gradient is assumed, the solution of Eq.12 is performed through a linear stability analysis. We start from an initial flow (i.e. a base state (BS) with $\vec{u}_{BS} = 0$) that represents a stationary state of the system. Then, we subject the base state to an infinitesimal perturbation for the various physical variables (i.e. velocity, temperature and pressure)

$$\begin{aligned} \vec{u} &= \vec{u}_{BS} + \delta \vec{U}_p \\ T &= T_{BS} + \delta T_p \\ P &= P_{BS} + \delta P_p \end{aligned} \qquad (14)$$

where $\partial_z P_{BS} = -g\rho_0 [1 + \alpha \partial_z T_{BS} z]$ and we obtain the equations of motion by neglecting terms $O(\delta^2)$. $\delta$ stands for an infinitesimal constant coefficient that is used to ensure that the additional term is a perturbation. Calculations and reorganization of the terms in the equations yield (for a temperature gradient along the $z$-axis)

$$\begin{aligned} \partial_t \vec{\nabla}^2 U_p^{(z)} &= \alpha g \left( \partial_x^2 + \partial_y^2 \right) T_p + (\mu/\rho_0) \vec{\nabla}^4 U_p^{(z)} \\ \partial_t T_p - \left( \partial_z T_{BS} \right) U_p^{(z)} &= K \vec{\nabla}^2 T_p \end{aligned} \qquad (15)$$

where the initial problem has been reduced to a coupled system of partial differential equations that contain the perturbation for the vertical (i.e. $z$-axis) component of the velocity, $U_p^{(z)}$, and the temperature $T_p$.

The form of the Eq.15 allows to introduce separable normal mode solutions $W(z)e^{-\omega t}\sum_n c_n e^{i\vec{k}_n \cdot \vec{r}}$ and $\theta(z)e^{-\omega t}\sum_n c_n e^{i\vec{k}_n \cdot \vec{r}}$ for the velocity (along $z$-direction) and the temperature component and solve the produced partial differential equations semi-analytically or numerically by applying appropriate boundary conditions depending on whether boundaries are no-slip or stress free [35, 36]. The solution allows to determine the condition for which $\omega<0$ and subsequently Marangoni or Rayleigh number as a function of a wavenumber $|k_R|$ that produces the lowest Marangoni or Rayleigh number at which a disturbance with this periodicity becomes unstable (see [35, 36]). The wavenumber that leads to the critical Marangoni or Rayleigh numbers constitutes the wavenumber that corresponds to the periodicity at the onset of the hydrothermal convection.

To describe the movement and convection of the fluid in the case that temperature gradient is not vertical, an appropriate modification is required to take into account that all thermo-physical quantities that appear in the heat transfer and Navier-Stokes equations have a spatio-temporal dependence. Given the variable depth of the crater, we introduce a local Cartesian coordinate system to describe movement on an inclined layer which is characterized by a variable angle $\gamma \equiv \gamma(x,y,z)$. A position in the new coordinate system is characterized by $(X' \equiv x, Y', Z')$ where the unit vectors are given by the expressions

$$\hat{Y}' = \cos(\gamma)\hat{y} + \sin(\gamma)\hat{z}$$
$$\hat{Z}' = -\sin(\gamma)\hat{y} + \cos(\gamma)\hat{z} \qquad (16)$$

where the original Cartesian coordinate system is characterised by the unit vectors $\hat{x}, \hat{y}, \hat{z}$. Assuming only steady motion the heat transfer equation yields

$$\frac{d^2 T_{BS}(X',Z')}{d(X')^2} + \frac{d^2 T_{BS}(X',Z')}{d(Z')^2} = 0 \qquad (17)$$

while the $X'$, $Z'$ of Eq.15 give (considering that $\vec{g} = g\left(-\sin(\gamma)\hat{Y}' + \cos(\gamma)\hat{Z}'\right)$ and $\vec{u}_{BS} \equiv \vec{u}_{BS}(x,Z')$)

$$\frac{1}{\rho_0}\frac{\partial P_{BS}(X',Y',Z')}{\partial X'} = (\mu/\rho_0)\left[\frac{d^2 u^{(x)}_{BS}(X',Z')}{d(X')^2} + \frac{d^2 u^{(x)}_{BS}}{d(Z')^2}\right]$$

$$\frac{1}{\rho_0}\frac{\partial P_{BS}(X',Y',Z')}{\partial Y'} = g\alpha T_{BS}(X',Z')\sin(\gamma) + (\mu/\rho_0)\left[\frac{d^2 u^{(y)}_{BS}(X',Z')}{d(X')^2} + \frac{d^2 u^{(y)}_{BS}}{d(Z')^2}\right] \qquad (18)$$

$$\frac{1}{\rho_0}\frac{\partial P_{BS}(X',Y',Z')}{\partial Z'} = g\alpha T_{BS}(X',Z')\cos(\gamma) + (\mu/\rho_0)\left[\frac{d^2 u^{(z)}_{BS}(X',Z')}{d(X')^2} + \frac{d^2 u^{(z)}_{BS}}{d(Z')^2}\right]$$

Solving numerically the above PDEs leads to the derivation of the spatial dependence of the base state for pressure, velocity and temperature. Then, we subject again as in the simpler case, the base state to an infinitesimal perturbation for the various physical variables (i.e. velocity, temperature and pressure)

$$\vec{u} = \vec{u}_{BS} + \delta \vec{U}_p$$
$$T = T_{BS} + \delta T_p \qquad (19)$$
$$P = P_{BS} + \delta P_p$$

to obtain the equations of motion by neglecting terms $O(\delta^2)$. $\delta$ stands for an infinitesimal constant coefficient that is used to ensure that the additional term is a perturbation. As previously, the perturbation of velocity, temperature and pressure will be introduced in the form of periodic functions (normal modes) and the system of the PDEs is solved to ensure that $\omega<0$. These conditions allow to determine the critical value of the Marangoni number and the associated wavenumber that yields instability. Fig.3 illustrates the (normalized to one) Marangoni number on the wavenumber $k_R$. It is evident that the minimum of the Marangoni number occurs at $|k_R| \sim 11\mu m^{-1}$ that corresponds to convection rolls with periodicity ~569nm.

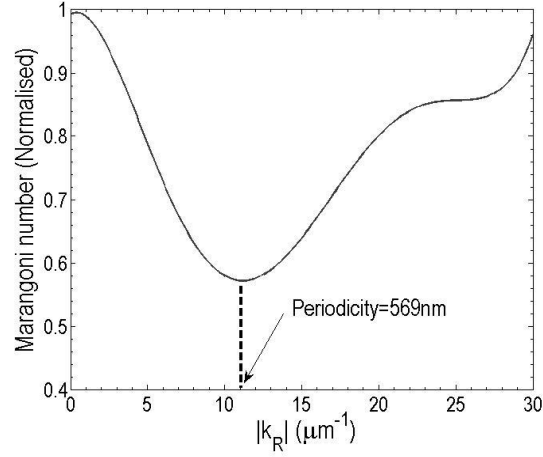

FIG. 3. Marginal stability of the hydrothermal patterns for $NP = 7$ occurs at $\left|k_R\right| \sim 11 \mu m^{-1}$ ($E_d$= 3.33 J/cm$^2$, $\tau_p$ = 170 fs, $\lambda_L$=513nm).

### D. Efficacy factor computation for $\lambda_L$=513nm and $\lambda_L$=1026nm.

The efficacy factor computation [32] yields the expected periodicity of the structures produced on the irradiated zone as a result of the excited electron density. To correlate the excited electron density with surface structures, the inhomogeneous energy deposition into the irradiated material is computed through the calculation of the product $\eta(k_L,k_i) \times |b(k_L)|$ as described in the Sipe-Drude model [32]. In the above expression, $\eta$ describes the efficacy with which the surface roughness at the wave vector $k_L$ (i.e. normalized wavevector $|k_L|=\lambda_L/\Lambda$) induces inhomogeneous radiation absorption, $k_i$ is the component of the wave vector of the incident beam on the material's surface plane and $b$ represents a measure of the amplitude of the surface roughness at $k_L$. To introduce surface roughness, the value $b$=0.4 is initially assumed [27]. Below, Fig.4 illustrates the computed efficacy factor at the normalised wavevector components $k_x$ (=$\lambda_L/\Lambda_x$), $k_y$ (=$\lambda_L/\Lambda_y$) for $\lambda_L$=1026nm (Fig.4a) and 513nm (Fig.4c), respectively. The values of the maximum carrier density used in the simulation are: $2.3 \times 10^{21}$cm$^{-3}$ and $7.5 \times 10^{21}$cm$^{-3}$, respectively, for the two wavelengths and they were derived by the solution of the MRE Eqs.1. According to Sipe theory, ripples are formed where the efficacy factors shows sharp points. A cross section across the $k_x$=0 shows that although for $\lambda_L$=1026nm a sharp minimum appears at $k_x$=1.1639 which corresponds to the formation of rippled structures of periodicity $\Lambda$~881nm (Fig.6b). This is in agreement with Fig.5a in the main manuscript. By contrast, no sharp points appear for $\lambda_L$=513nm (Fig.6d). This suggests that no parallel periodical structures will be produced in the region characterised by large electron densities (see Fig.2a in the main manuscript). Furthermore, although horizontal (i.e. along the $k_x$ axis which are also characterised by sharp points, they correspond to non-physical modes that is perpendicular ripples, as they are not observed).

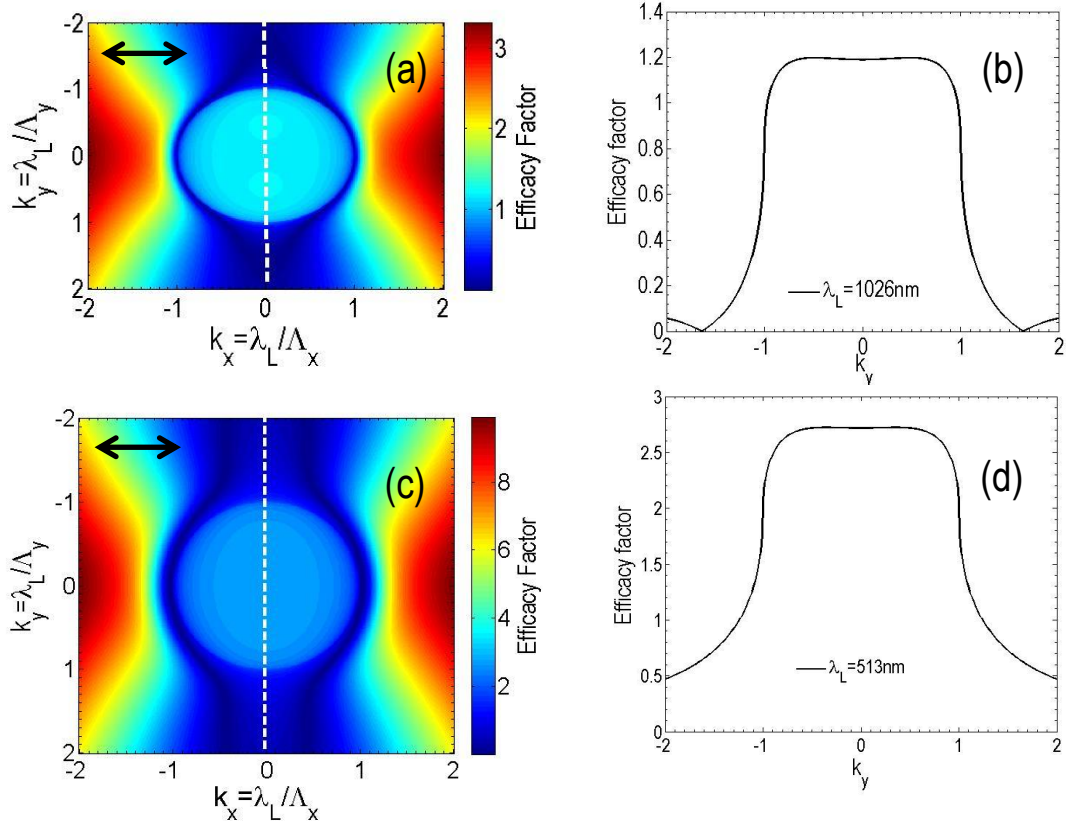

FIG. 4. Efficacy factor distribution at the normalised wavevectors $k_x$, $k_y$ for $n_e=2.3\times10^{21}\text{cm}^{-3}$ and $n_e=7.5\times10^{21}\text{cm}^{-3}$ for $\lambda_L=1026$nm (a) and $\lambda_L=513$nm (c), respectively. (b), and (d) illustrate cross section along the *white* dotted line ($\tau_p=170$fs, $E_d=3.33/\text{cm}^2$, the double-ended arrows indicate the laser beam polarization).

### E. Structure periodicities.

The following graphs provide the simulation results for surface profile (i.e. a quadrant) irradiated with *NP*=6 (Fig.5a) and *NP*=12 (Fig.5c) ($E_d=3.33\text{J/cm}^2$ and $\lambda_L=513$nm). The height change along the *y*-axis (*white* dashed line in Figs. 5a,c) are sketched in Figs.5b,d, respectively. It is evident that the empty space inside the $r<r_G$ (i.e. $r_G=8\mu$m) region for *NP*=6, is then populated with supra-wavelength grooves. By contrast, for $\lambda_L=513$nm, conventional ripples are formed everywhere in the quadrant (Fig.6a,b).

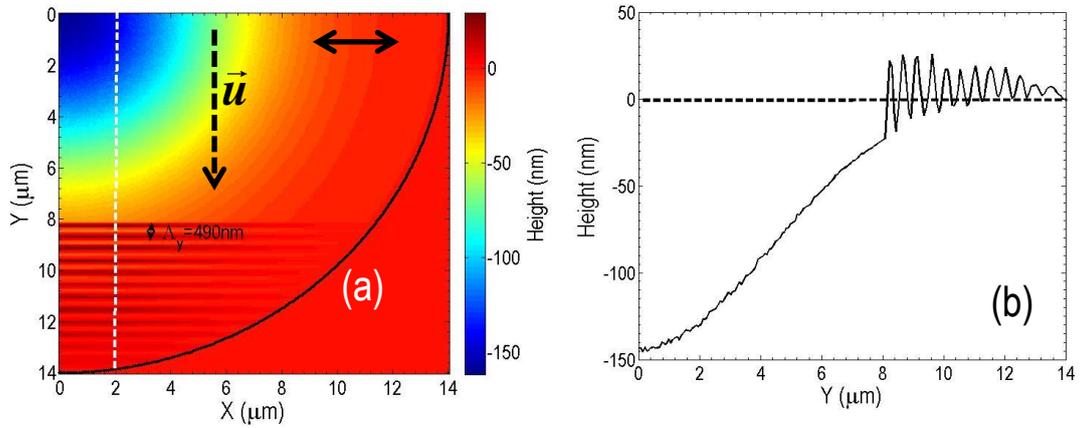

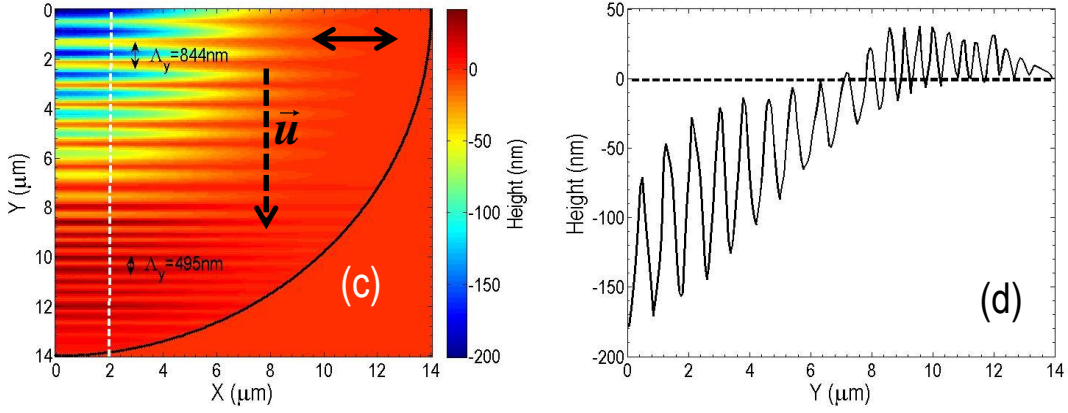

FIG.5. Surface profile (quadrant) for $\lambda_L$=513nm. (a) Spatial profile (*NP*=6, ripple periodicity: $\Lambda_y$~490nm). (b) Height change along the *Y*-axis at *X*=2μm which is indicated by the white dashed line in (a). (c) Spatial profile (*NP*=12, ripple periodicity at ~495nm, groove periodicity at ~844nm). (d) Height change along the *X*-axis which is indicated by the white dashed line in (c) ($E_d$=3.33/cm$^2$, the double-ended arrow indicate the laser beam polarization/black line in (a,c) indicates the border of the affected zone). The *dashed* lines in (a), (c) indicate the preferred movement of the material in liquid phase ($\vec{u}$ is the tangent velocity field of the molten profile).

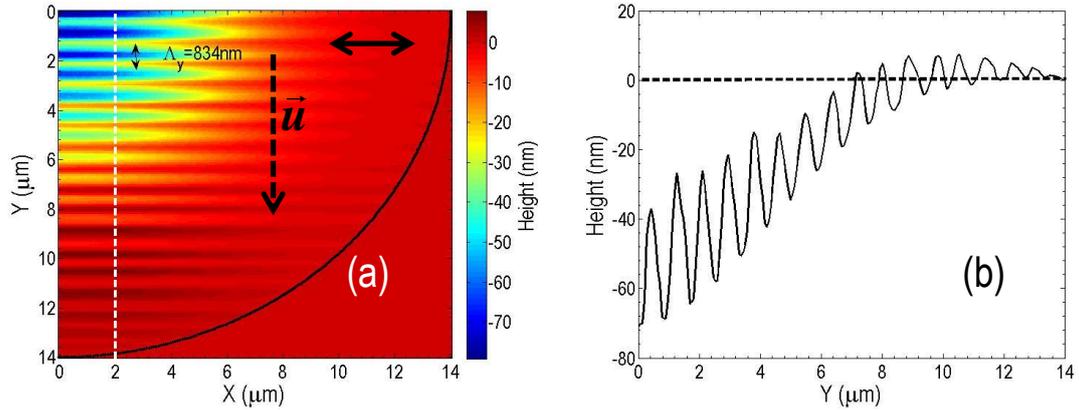

FIG.6. Surface profile (quadrant) for *NP*=4, $\lambda_L$=1026nm. (a) Spatial profile (ripple periodicity $\Lambda_y$=834nm at ~$\lambda_L/n$). (b) Height change along the *Y*-axis at *X*=2μm which is indicated by the white dashed line in (a). ($E_d$=3.33/cm$^2$, the double-ended arrow indicate the laser beam polarization/black line in (a) indicates the border of the affected zone). The *dashed* line in (a) indicates the preferred movement of the material in liquid phase ($\vec{u}$ is the tangent velocity field of the molten profile).

## F. Experimental results.

The periodic structures formed after the irradiation of the surface were characterized by Scanning Electron Microscopy (JEOL JSM-7500F). SEM images were then analysed using a two-dimensional Fast Fourier Transform (2D-FFT) to calculate the periodicity as shown in Fig.7. The 2D-FFT image of the original images (Fig.7a) reveals the direction of the periodicity which is horizontally and so the profile of the image was taken appropriately as shown in Fig.7b. The distance between the central peak and peak 1 or 2 of Fig.7c represent the frequency *f* of the periodic structure. In order to calculate the periodicity, *Λ*, of the structures first we calculate the average frequency of *c1* and *c2*, and so the average period is *<Λ>=1/f*. The error of every analysis is calculated as follows

$$\Delta\Lambda = \left| -\frac{1}{f^2} \right| \cdot \Delta f$$

The calculated average period is (888.5± 126.1) nm.

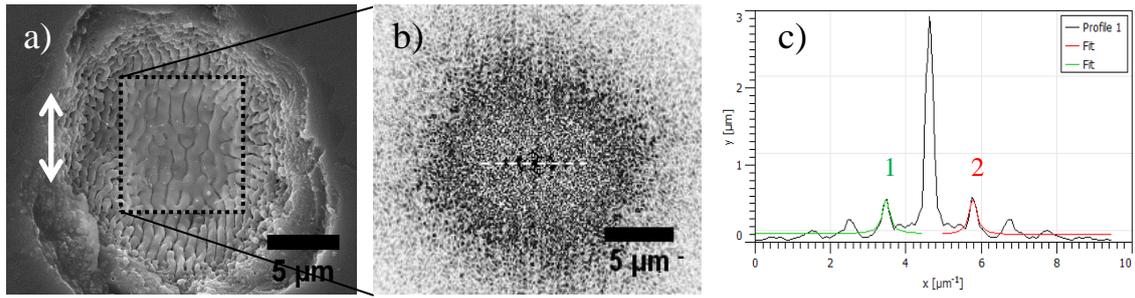

FIG. 7. (a) SEM images (top view) on fused silica surface after irradiation with 1.85J/cm$^2$ at 513nm for *NP*=20. (b) 2D-FFT image of the selected area. (c) Profile of the white dashed line from image (b) (the double-ended arrow indicates the laser beam polarization).

After the irradiation of fused silica surface with ultrafast femtosecond laser pulses, periodic structures are observed such as ripples and grooves. Both of those structures are parallel to the polarization of the incident beam. As explained in the main manuscript and the above paragraphs, the enhanced energy of photons corresponding to laser beam at smaller wavelengths allows the depletion of a region of periodical structures (Fig.8a) with a population with grooves and ripples at higher number of pulses (Fig.8b). By contrast, at larger wavelengths, the excitation levels are not sufficient enough to deplete part of the affected zone for small *NP* (Fig.8c) while at higher NP, the ripples still exist without formation of structures with larger periodicity (Fig.8d).

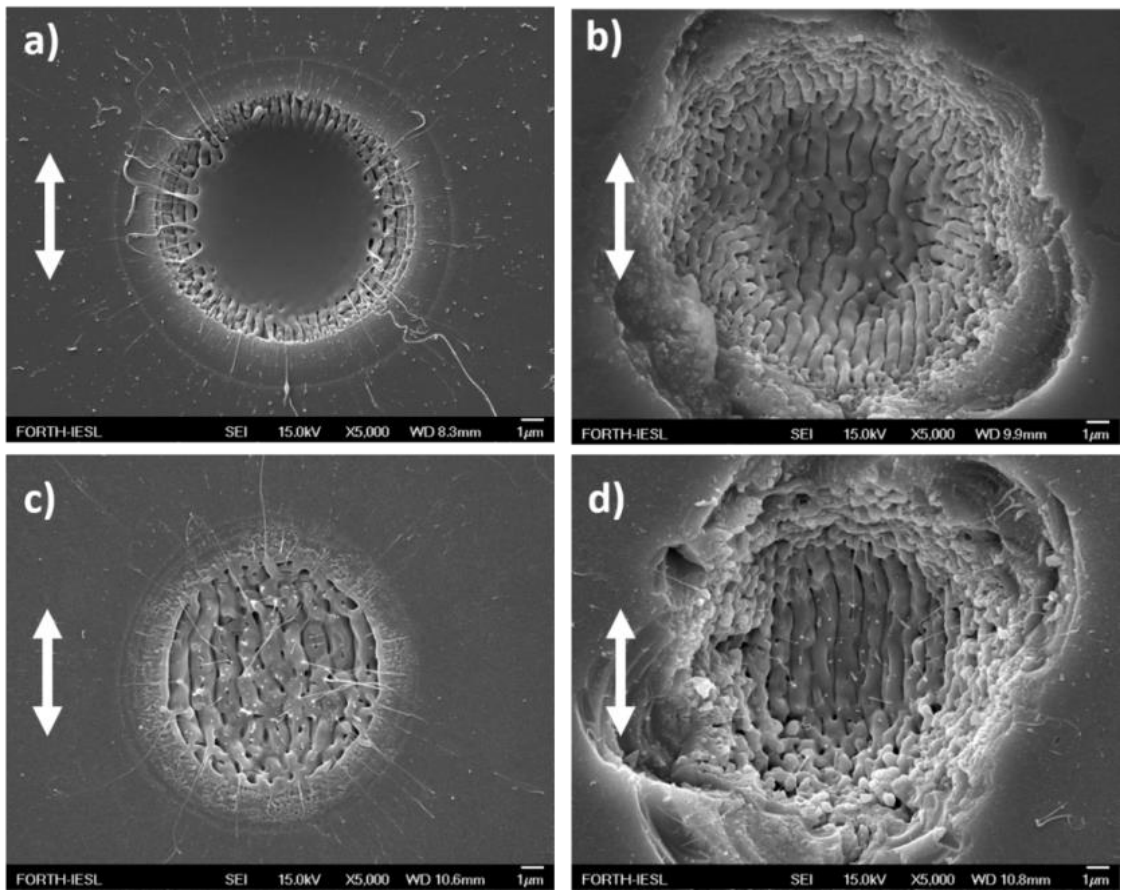

FIG. 8. (SEM images (top view) on fused silica surface after irradiation with 1.85J/cm$^2$ at 513nm and 1026nm for *NP*=4 (a)-(b) and *NP*=20 pulses (c)-(d), respectively. (The double-ended arrow indicates the laser beam polarization).

The same experimental procedure was conducted for crystalline silicon oxide (Quartz) where the morphological results showed similar behavior with fused silica regarding the orientation and the periodicity of periodic structures.

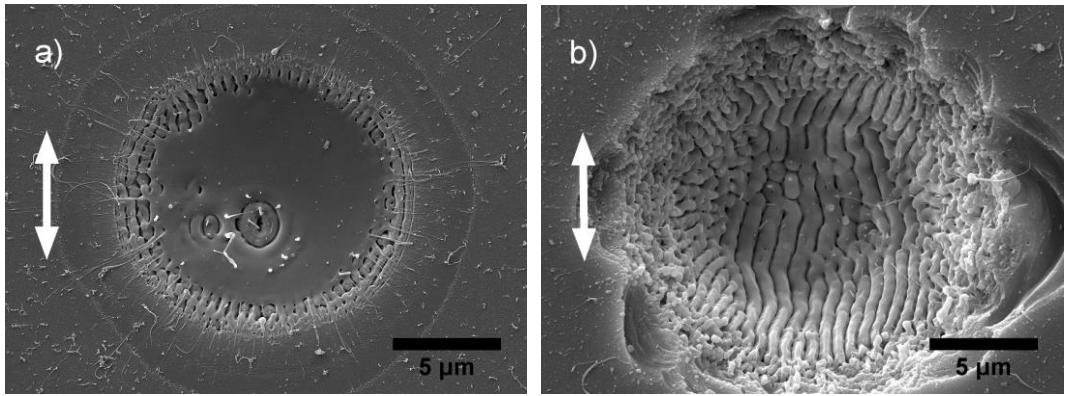

FIG. 9. (SEM images (top view) on quartz surface after irradiation with 3.33J/cm$^2$ at 513nm for *NP*=4 (a) and *NP*=20 pulses (b). (The double-ended arrow indicates the laser beam polarization).